\documentclass[twocolumn,showpacs,preprintnumbers,prl,fleqn,floatfix]{revtex4}

\usepackage{amssymb}
\usepackage{epsfig}
\usepackage{amsmath}

\begin{document}

\title{Optical absorption to probe the quantum Hall ferromagnet at filling factor $\nu=1$}
\author{P. \surname{Plochocka}}
\email{Paulina.Plochocka@grenoble.cnrs.fr}
\author{J. M. \surname{Schneider}}
\author{D. K. \surname{Maude}}
\author{M. \surname{Potemski}}
\affiliation{$^{}$Laboratoire National des Champs Magn\'etiques Intenses, Grenoble High Magnetic Field Laboratory,
CNRS, 38042 Grenoble, France }

\author{M. \surname{Rappaport}}
\author{V. \surname{Umansky}}
\author{I. \surname{Bar-Joseph}}
\affiliation{ Department of Condensed Matter Physics, The Weizmann
Institute of Science, Rehovot, Israel}

\author{J. G. \surname{Groshaus}}
\author{Y. \surname{Gallais}}
\altaffiliation[Current address: ]{LMPQ, Universit\'e Paris Diderot, 75205 Paris, France}
\author{A. \surname{Pinczuk}}
\affiliation{$^{}$ $^{}$Department of of Physics and of Appl.
Physics and Appl. Mathematics, Columbia University, New York, NY
10027}

\date{\today }

\begin{abstract}
Optical absorption measurements are used to probe the spin polarization in the integer and fractional quantum Hall
effect regimes. The system is fully spin polarized only at filling factor $\nu=1$ and at very low temperatures
($\sim40$~mK).  A small change in filling factor ($\delta\nu\approx\pm0.01$) leads to a significant depolarization.
This suggests that the itinerant quantum Hall ferromagnet at $\nu=1$ is surprisingly fragile against increasing
temperature, or against small changes in filling factor.
\end{abstract}

\maketitle

Electron-electron interactions in two dimensions dominate in many cases over the single particle physics leading to new
collective ground states of the system. This is particularly true in GaAs due to the small value of the single particle
Zeeman energy. The physics in the vicinity of filling factor $\nu=1$ is particularly rich. The system behaves as a half
empty Landau band in which all the electrons have the same orbital quantum number and only the spin degree of freedom
remains. At exactly $\nu=1$, the predicted ground state is an itinerant quantum Hall ferromagnet
\cite{Kasner96,Jungwirth}, while on either side of $\nu=1$ the system depolarizes more rapidly than predicted by the
single particle picture, due to the formation of spin textures (Skyrmions or anti-Skyrmions) in the ground state
\cite{Sondhi,Fertig,Barrett,Schmeller95,Maude96,Aifer96}. The strong coupling between the nuclear and electronic spin
systems, observed close to $\nu=1$ in specific heat capacity~\cite{Bayot96}, and resistively detected nuclear magnetic
resonance (NMR) measurements~\cite{Desrat02}, strongly suggest the existence of gapless spin excitations of the
electronic system. Such Goldstone modes are consistent with a breaking of the spin rotational symmetry due to the
formation of a Skyrme crystal in the ground state~\cite{Brey}.

Electrical transport measurements, which have been extensively used to investigate the quantum Hall effect, are not an
incisive probe of the physics of the ground state at exactly integer filling factor, since the Fermi energy lies deeply
inside localized states. Optical techniques such as photoluminescence, absorption and inelastic light scattering have
been widely applied \cite{Aifer96,Kukushkin,Chughtai02,Plochocka2007,gallais08,Zhuravlev08}. Surprisingly, techniques
which give a \emph{direct} measure of the spin polarization, suggest that the system is not fully spin polarized at
$\nu=1$~\cite{Aifer96,Chughtai02,Zhuravlev08}, despite the large exchange enhanced spin gap which remains open even in
the absence of the single particle Zeeman energy~\cite{Maude96}.

In this paper we report on optical absorption (transmission) measurements to directly probe the subtle physics of the
n=0 Landau level (LL) via the spin polarization of the system. We find that full spin polarization does indeed occur,
but only at exactly filling factor $\nu=1$ and at very low (40~mK) temperatures. This suggest that the quantum Hall
ferromagnet at $\nu=1$ is surprisingly fragile, collapsing, with either a small change of filling factor or
temperature, into a lower energy ground state with a large number of reversed spins.

To measure the absorption spectrum of a \emph{single} GaAs quantum well (QW) at low temperatures we have used a
structure which forms a half-cavity for the incoming light with the QW located at the anti-node of the standing wave
formed by the optical field~\cite{Plochocka2007}. The sample is illuminated by white light and the reflected spectrum
is measured. The modulation doped 20nm wide GaAs QW and Bragg mirror are grown on an $n^{+}$ layer that serves as a
back gate. The wafer was processed to form a mesa structure, with ohmic contacts to the QW and to the back gate, such
that $n_{e}$ can be tuned in the range $(0.4-3)\times 10^{11} $ cm$^{-2}$. The measured electron mobility is $\sim
1\times 10^{6}$cm$^{2}$ V$^{-1}$ s$^{-1}$.

The magnetic field is applied parallel to the growth direction. Reversing the magnetic field (keeping the circular
polarizer fixed) corresponds to changing the circular polarization of the light: at positive fields we detect
transitions from the heavy hole band to the lower electron Zeeman spin subband (LZ) and at negative field - to the
upper electron Zeeman spin subband (UZ). We label the transition to the LZ as $\sigma ^{+}$, and to the UZ as $\sigma
^{-}$. Simultaneously with the absorption measurements the two terminal transport data have been taken in an effective
Corbino geometry using an \emph{ac} excitation current of 0.1~nA at 10.7~Hz, and phase sensitive detection. The
measured \emph{ac} voltage is proportional to $1/\sigma_{xx}(B)$. In this experiment the value of the filling factor
can be changed either by tuning the magnetic field or the gate voltage.

\begin{figure}[]
\begin{center}
\includegraphics[width=0.96\linewidth]{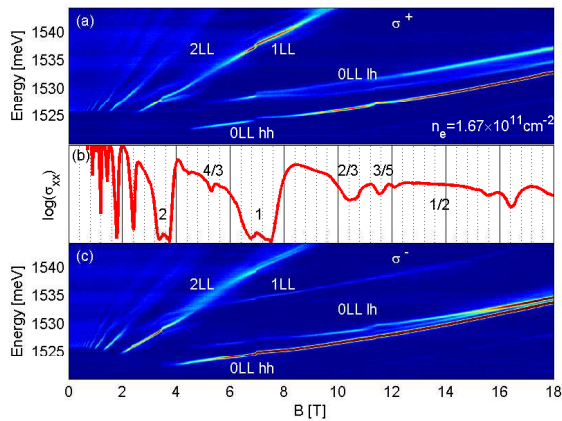}
\end{center}
\caption{(color on-line) (a) and (c) the $\sigma ^{+}$\ and $\sigma ^{-}$\ optical density spectra (minus logarithm of
the normalized transmission) as a function of the magnetic field for $n_{e}=1.67 \times 10^{11}$ cm$^{-2}$ and
$T=40$~mK. (b) simultaneously measured electrical transport showing logarithm of the conductivity versus magnetic
field.} \label{fig1}
\end{figure}

Typical results of the measurements  taken as a function of magnetic field for a carrier density $n_{e}=1.67 \times
10^{11}$ cm$^{-2}$ are presented in Fig~\ref{fig1} where we plot the optical density (minus logarithm of the normalized
transmission) versus magnetic field. A compilation of absorption spectra for $\sigma ^{+}$, $\sigma ^{-}$ polarization
is shown in Fig~\ref{fig1}(a),(c) respectively. Figure~\ref{fig1} (b) shows the results of the simultaneous transport
measurements of $\sigma_{xx}(B)$.

Before focusing on the n=0 LL, we start with an overview of the data taken as a function of the magnetic field
(Fig~\ref{fig1}). In the presented spectra the absorption to a given (n-th) electronic LL is observed. One can
distinguish the transitions for LL $n=6$ to $n=1$. The energy of the transitions of these levels follows nicely the
$(n+1/2) \hbar \omega_{c}$ linear dependence with the magnetic field.

The absorption to a given LL is strong over a limited range of the magnetic field. The appearance of absorption to a
given LL above a critical magnetic field can be explained by a band filling argument: Absorption to the n-th LL is
impossible while ever it is full, but becomes allowed once it starts to depopulate due to the increasing $eB/h$ LL
degeneracy. With increasing magnetic field the LL progressively empties resulting in an increase in the absorption. For
sufficiently large magnetic field, the LL is completely empty, and the absorption decreases, even though absorption
remains allowed to the empty level. As discussed in reference~\cite{Plochocka2007}, this effect can be attributed to
the Fermi edge singularity.

The magnetic field at which the Fermi energy jumps down to a given LL (onset of the absorption to a given LL) allows us
to assign a filling factor, and hence determine the carrier density. The optical spectra are fully consistent with the
filling factors determined from the transport data in Fig.~\ref{fig1}(b). Minima in $\sigma_{xx}(B)$ which occur for
odd and even filling factors are in good agreement with the onset of absorption to a given LL in the optical spectra.

The physics of the $n=0$ LL is clearly very different. The absorption starts at a magnetic field corresponding to
filling factor $\nu=2$ \emph{for both} circular polarizations of the light (Fig~\ref{fig1}(a,c)). Thus the LZ level is
never fully occupied between $\nu=2$ and $\nu=1$ in striking contradiction to the prediction of the single particle
model. At $\nu=1$ we observe a minimum of the absorption for $\sigma ^{+}$ and maximum for $\sigma ^{-}$ polarization.
At certain filling factors such as $\nu=1, \nu=3/5$, a significant change in the position of the absorption line is
observed. This kind of behavior has been predicted, and, observed in photoluminescence experiment for integer filling
factors \cite{Hawrylak97,Gravier98} as well as for fractional ones \cite{Yusa01, Wojs00, Byszewski06}. It was suggested
that the shift of the line was the result of screening by the sea of the electrons which is modified for a given
fractional or integer (incompressible) quantum Hall state.

\begin{figure*}[t]
\begin{center}
\includegraphics[width=0.9\linewidth]{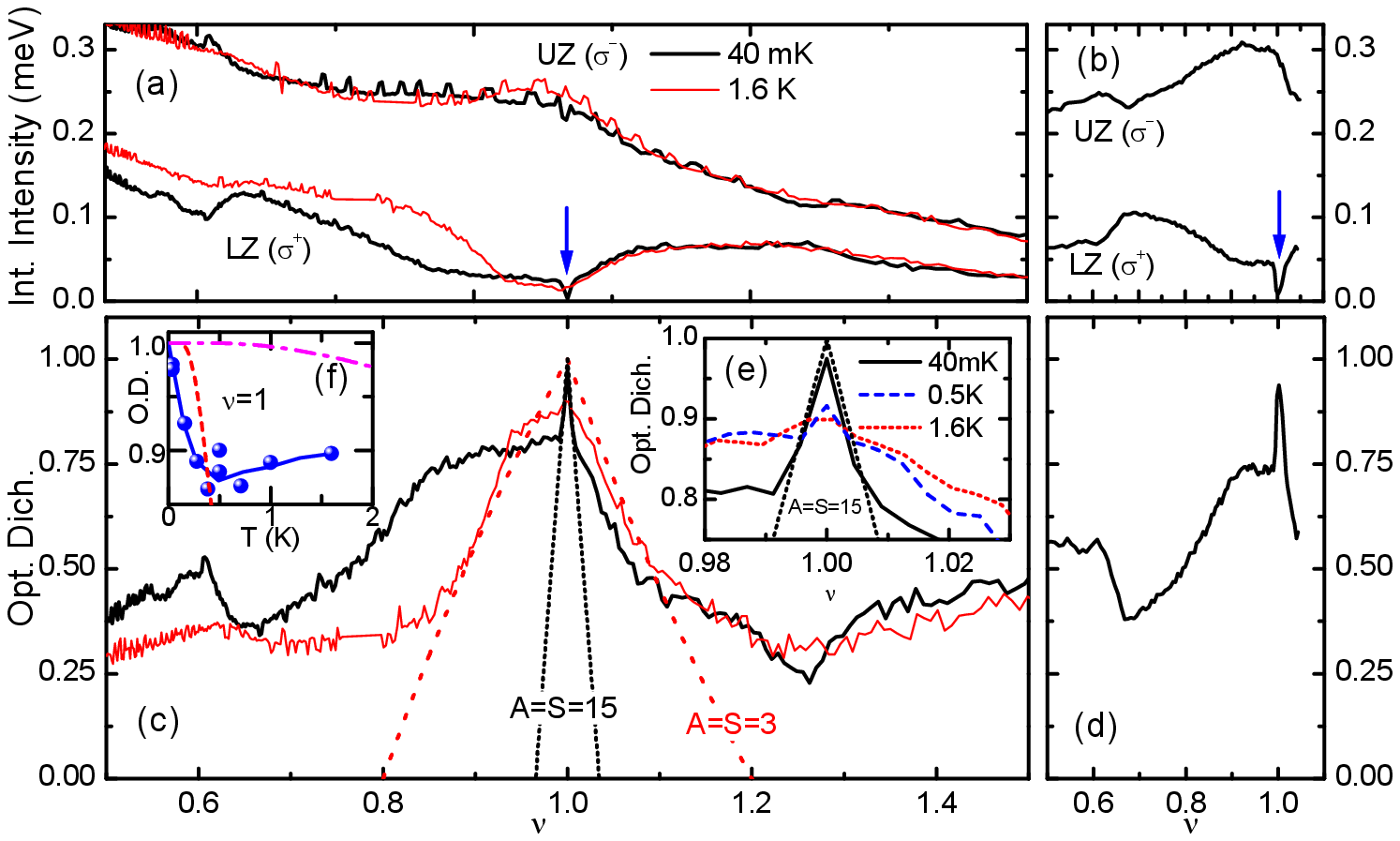}
\end{center}
\caption{(color on-line) Integrated intensity ($I_{\sigma^{\pm}}$) of the absorption to the $n=0$ LL measured for both
$\sigma ^{+}$\ and $\sigma ^{-}$ polarizations as a function of filling factor for (a) a constant density of electrons
$n_{e}=1.67 \times 10^{11}$ cm$^{-2}$ and (b) a constant magnetic field $B=11$~T. (c), (d) optical dichroism (spin
polarization) calculated using the data presented in (a) and (b). The calculated depolarization for finite size
skyrmions/anti-skyrmions is shown by the dashed (A=S=3) and dotted (A=S=15) lines. (e) shows the spin polarization
around filling factor $\nu=1$ measured at T=40, 500~mK and 1.6~K. (f) shows the detailed temperature dependence of the
spin polarization at exactly $\nu=1$ (the solid line is drawn as a guide to the eye). The broken lines are the
predicted temperature dependence of the polarization for a spin wave excitation (dash-dot) and for two spin levels
separated by the single particle Zeeman energy (dashed) as described in the text.} \label{fig2}
\end{figure*}

The oscillator strength can be obtained by integrating the optical density of a given absorption line as a function of
the magnetic field or the applied gate voltage. Typical results for the oscillator strength for absorption to the $n=0$
LL are presented at Fig~\ref{fig2}(a,b). The data presented in Fig~\ref{fig2}(a) show the integrated intensity of the
$n=0$ LL for both circular polarizations measured as a function of the magnetic field (plotted here as a function of
filling factor). Data is shown for two different temperatures: $40$mK and $1.6$K. The integrated intensity measured at
$40$mK and constant magnetic field, as a function of the gate applied to the sample (plotted here as a function of
filling factor), is also presented in Fig~\ref{fig2}(b). For the data taken at low temperature a very similar behavior
of the integrated intensity as a function of filling factor is observed. For both polarizations ($\sigma ^{+}$ and
$\sigma ^{-}$) the absorption is non zero over almost all the range $0<\nu<2$. This can also be directly seen in the
raw data in Fig~\ref{fig1}. The most important observations are: (i) for the absorption to the LZ ($\sigma ^{+}$) a
sharp minimum at $\nu=1$ is observed while at lower filling factors the integrated intensity forms a plateau down to
$\nu\cong 0.85$. (ii) at $\nu=2/3$ well developed maximum of the absorption to LZ is observed which corresponds to the
minimum for absorption to UZ ($\sigma ^{-}$). The latter is even more pronounced when the filling factor is changed by
applying the gate (Fig.\ref{fig2}(b)). This is due to the fact that as we decrease the filling factor by increasing the
magnetic field simultaneously the absorption increases because of the increasing $eB/h$ LL degeneracy.

The fact that the absorption to LZ is non zero above filling factor $\nu=1$ is experimental evidence that the LZ is not
fully occupied due to additional population of the UZ. A zero of the absorption at exactly $\nu=1$ (LZ transition)
shows that LZ is fully occupied so that absorption is not possible. On the other hand, for absorption to UZ the maximum
of the integrated intensity is observed when the absorption takes place to an empty level. As the temperature increases
(see the 1.6K data in Fig~\ref{fig2}(a)) the sharp transition to zero absorption to LZ at $\nu=1$ is no longer
observed. In addition, the changes of the integrated intensity characteristic for $\nu=2/3$ are no longer visible as
one could expect for fractional states.

Absorption measurements give direct access to the spin polarization of the 2DEG. The optical dichroism can be
calculated directly from the absorption and reflectivity measurements \cite{Aifer96, Manfra96, Chughtai02}. It is
however important to distinguish between the charged and neutral exciton absorption lines in the spectra
\cite{Groshaus2007}. A weaker absorption to the neutral exciton can be observed for lower carrier densities. For the
charged exciton line, the optical dichroism, $(I_{\sigma^-}-I_{\sigma^+})/(I_{\sigma^-}+I_{\sigma^+})$, is exactly
equal to the spin polarization of the 2DEG~\cite{Groshaus2007}. We have calculated the optical dichroism from the data
presented at Fig~\ref{fig2} (a) and (b). The results are shown in Fig~\ref{fig2}(c) and (d) respectively.  At low
temperature the optical dichroism (spin polarization) measured either for constant density of electrons or constant
magnetic field is very similar. First of all, neglecting the sharp feature at $\nu=1$ the data are symmetric around
$\nu=1$. Notably, well developed minimum at $\nu=2/3$ and $\nu=4/3$ are seen. This symmetry, previously reported both
in transport \cite{Du95} and reflectivity \cite{Chughtai02} measurements, was discussed in terms of particle-hole
symmetry; unoccupied states in the $n=0$ LL for filling factors $1<\nu <2$ can be treated as holes with an effective
filling factor $\nu_{h}=2-\nu_{e}$.

The measured polarization at $T=1.6$~K is comparable with previous absorption and reflectivity measurements
\cite{Aifer96, Chughtai02}. The spin polarization saturates at approximately 0.8 and the depolarization on both sides
of $\nu=1$ is roughly symmetric and compatible with the formation of spin textures (Skyrmions or anti-Skyrmions) in the
ground state of size $S\approx A \approx 3$ (dashed lines in Fig.\ref{fig2}(c)), as previously reported for samples of
similar density \cite{Aifer96}. At fractional filling factors $\nu=2/3$ and $4/3$, and mK temperatures, the
polarization shows minima, although the system never fully depolarizes (or polarizes), in contrast to the expectation
for integer composite Fermion filling factors, but predicted theoretically from numerical studies of finite size
systems~\cite{Vyborny07}. What is new in our data, is that at very low temperature ($T=40$~mK), the system does indeed
fully polarize within experimental error ($\sim0.97$) at exactly filling factor $\nu=1$, and that this feature is
extremely sharp. This can be seen more clearly in the right insert of Fig.\ref{fig2}(c) which shows an expanded view
around $\nu=1$. The width of this feature (FWHM) is only $\delta \nu\approx0.01$ and was only seen at all because of
the very precise measurements made. This feature is a direct consequence of the sharp dip to zero in the absorption
observed for the transition to the LZ (indicated by arrows in Fig.~\ref{fig2}(a-b)). A zero value of the absorption at
$\nu=1$, a prerequisite for full polarization, is not presented convincingly in Ref.~\cite{Manfra96}.

The fully polarized state is a signature of the quantum Hall ferromagnet at filling factor $\nu=1$
\cite{Kasner96,Jungwirth}. The sharp depolarization observed either side of $\nu=1$ corresponds to $\approx15$ spin
flips per magnetic flux quanta added or removed from the system (see short dotted lines in Fig.\ref{fig2}(c)). This
result is consistent with the formation of a lower energy spin texture (skyrmion or anti-skyrmion) ground state either
side of $\nu=1$ or with the formation of spin reversed domains. The presence of spin textures close to $\nu=1$ is
supported by evidence for gapless excitations in the spin excitation spectra using specific heat, resistively detected
NMR and inelastic light scattering~\cite{Bayot96,Desrat02,gallais08}, and attributed to the presence of a Skyrme
crystal.

The detailed temperature dependence of the spin polarization at exactly $\nu=1$ can be seen in the Fig.\ref{fig2}(f). A
temperature of a few hundred mK is already sufficient to suppress full spin polarization. The thermodynamics of the
$\nu=1$ quantum Hall ferromagnet should be governed by thermal activation to the continuum of the spin-exciton (spin
wave) excitation spectrum~\cite{Jungwirth,Kasner96} so that the low temperature spin polarization is proportional to
$1+C k_{B}T\ln(1-e^{g\mu_{B}B/k_{B}T})$. This prediction is shown by the dot-dashed line in Fig.~\ref{fig2}(f) and, as
can be seen, largely overestimates the robustness of the $\nu=1$ quantum Hall ferromagnet observed in our experiments.
The NMR probed $\nu=1$ ferromagnet~\cite{Barrett} remains more fragile to temperature than predicted by the
spin-exciton model, but is still far more robust than the behavior reported here. However, in the NMR measurements,
full spin polarization was arbitrarily attributed to the measured value of the Knight shift at $\nu=1$ at $T=1.5$~K and
in addition the temperature dependence was measured for $\nu=0.98$. It seems therefore likely, that full spin
polarization was not achieved under the measurements conditions of ref.\cite{Barrett}. Surprisingly, the predicted spin
polarization calculated for two spin levels separated by the bare Zeeman energy, $g\mu_{B}B\simeq2.1$~K, plotted as a
dashed line in Fig.\ref{fig2}(f) coincides much better with our data. Therefore, our observations are more in line with
the recent conclusion of Zhuravlev and co-workers~\cite{Zhuravlev08} who report that $\nu=1$ ferromagnetism with long
range order is very sensitive to temperature and only persists when $k_{B}T< g\mu_{B}B$.

In conclusion, optical absorption measurements have been used to probe the spin polarization in the quantum Hall effect
regime. The system is fully spin polarized only at exactly filling factor $\nu=1$ and at very low temperatures
($\sim40$~mK). A small change in filling factor ($\delta\nu\approx\pm0.01$) leads to a significant depolarization of
the system. This, together with the temperature dependence, suggest that the itinerant quantum Hall ferromagnet at
$\nu=1$ is not robust, and collapses, whenever possible, to a lower energy ground state with a large number of reversed
spins.

\acknowledgements{One of us (PP) is financially supported by the European Community under FP7/2007-2013, contract no.
221249 `SESAM'. The work at LNCMI was partially supported by the European 6th Framework Program under contract number
RITA-CT-3003-505474.}


\end{document}